# Recent Advances of Nanomaterials in Membranes for Osmotic Energy Harvesting by Pressure Retarded Osmosis

**Arvin Shadravan, Mahmood Amani**

*Abstract*— Energy and water issues are the two main global challenges faced by the human in the past decade. The rapid growths in global energy consumption and greenhouse gas emissions have encouraged the exploration of renewable energy sources as substitute fuels. Osmotic energy (or salinity-gradient energy) is the energy released when water with different salinities is mixed, such as rivers and oceans. By employing a semipermeable membrane to control the mixing process, the osmotic pressure gradient energy can be generated in terms of electrical power via pressure retarded osmosis (PRO) without causing adverse environmental impacts. This work presents a review of the fabrication of thin film nanocomposite (TFN) membranes which are customized to offer high flux in forward osmosis (FO) and high osmotic power in PRO. The hydraulic pressure during PRO processes is less than RO processes so membranes that are used for PRO are less likely to foul. The application of these nanomaterials incorporated with TFN membranes for power generation through PRO is still a new field to be explored. Despite some promising findings obtained from this work, there is always room for improvement.

## I. INTRODUCTION

Osmotic energy (or salinity-gradient energy) is the energy released when water with different salinities is mixed, like rivers and oceans. Besides, pressure retarded osmosis (PRO) has improved to have a promising role as an approach for energy extraction [1]. The most efficient advantage of producing energy by the PRO process is that it can be continuous and non-stop and provide constant energy production. Although there are different kinds of methods that are available for energy generation through salinity gradients which contain PRO and RED, the most power generation was harvested by the PRO process [2]. If two solutions with various sorts of molecular structures such as seawater and river water connect, they will mix instantly and produce energy. Therefore, the concentration of salt is prone to produce energy. Osmotic pressure has been widely used for harvesting energy among renewable energies. In the following, free Gibbs energy of mixing is one of the ways to produce energy [3].

Arvin Shadravan, Texas A&M University, College Station, Texas, USA, +13463704425.

Mahmood Amani, Texas A&M University at Qatar, Doha, Qatar, +97455837368.

### A. Energy Generation through Salinity Gradient

Osmotic energy (or salinity-gradient energy) is the energy released when water with different salinities is mixed, like rivers and oceans. Besides, pressure retarded osmosis (PRO) has improved to have a promising role as an approach for energy extraction [1]. The most efficient advantage of producing energy by the PRO process is that it can be continuous and non-stop and provide constant energy production. Although there are different kinds of methods that are available for energy generation through salinity gradients which contain PRO and RED, the most power generation was harvested by the PRO process [2]. If two solutions with various sorts of molecular structures such as seawater and river water connect, they will mix instantly and produce energy. Therefore, the concentration of salt is prone to produce energy. Osmotic pressure has been widely used for harvesting energy among renewable energies. In the following, free Gibbs energy of mixing is one of the ways to produce energy [3].

(eq.1):
ΔG mix = Frictional Losses + Unutilized energy + Extractable Work

Osmotic energy will be produced after the process of mixing feed solution and draw solution with each other [4]. PRO has two significant parts that are PRO membrane module and the turbine by which can generate power [5]. Generally, the structure of PRO and RO membranes are not different from each other but the porosity in PRO membranes is more than RO membranes. Figure 1 shows the schematic of this process [6].

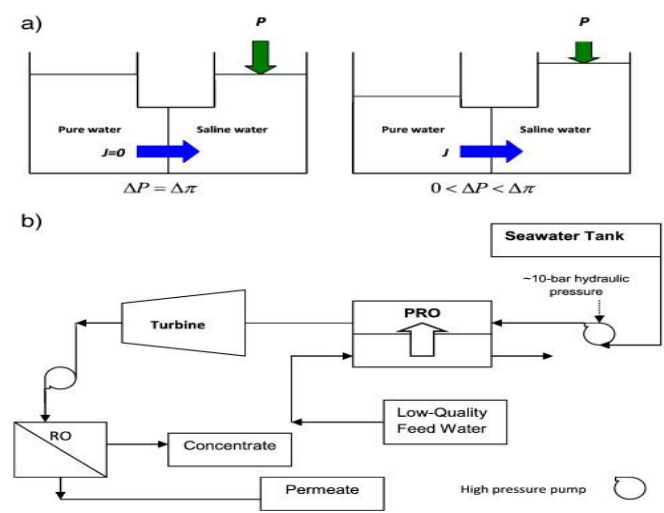

Figure 1: PRO a) the procedure perception b) PRO-RO

strategy [6]

When fresh water left the membrane and passes through the turbine to generate electricity then it will go back to the sea or preserved with a membrane or thermal functionality for reproduction [7].

Using seawater in the PRO process has some advantages and disadvantages. The first important benefit that it has, is the availability of seawater and the second one is having the free high osmotic pressure. The disadvantages are known as pumping and the pretreatment [6]. To calculate the water flux through the PRO process we need to calculate $J_W$ which is the sign of water flux and is shown by the following equation [8]: (eq.2):

$$J_W = A_W (\Delta\pi - \Delta P)$$

### B. Pressure Retarded Osmosis

During the period of the osmosis process, permeate water rapidly diluted draw solution, when the feed solution is being concentrated. The hydrostatic pressure is entitled to osmotic pressure such as water transportation passing through the membrane while it is functionalized to draw solution. At any step, while $\Delta P$ can be between zero to $\Delta\pi$, water can flow into the saline water because $\Delta\pi$ (Osmotic pressure) is still larger than the pressure which is $\Delta P$. This process is entitled to PRO [10]. Pressure retarded osmosis (PRO) was conceptualized for producing salinity gradient energy by [11] in the 1950s and then reinvestigated in the mid1970s due to the world's energy crisis [12]. Loeb and his coworkers conducted theoretical and experimental research on the feasibility of PRO [13].

### C. Benefits, Drawbacks and Challenges in PRO

It was predicted that all the potential energy sources can harvest almost 2000 TWh every year of power generation that can be more than 10% for the existing demand for energy in the world. The essential requirements in the membrane for the PRO process are: i) high water flux, ii) moderate tolerance to pH change and chlorine attack, iii) high mechanical strength and stability and iv) low internal concentration polarization effect. Hence, there are several highlights for the significant osmotic process through PRO technology which are as follows: a) energy generation through salinity gradient, b) basic PRO principles and performance interrelated with the highlights of power density, c) novel and promising PRO membranes, d) viability and perspectives about the future of the PRO process [14].

Osmotic potential is the main driving force to produce power density under the PRO process. The other challenging part of the PRO process is while increasing water flux (water permeability), the salt permeability will increase too. So finding the optimum point in which water flux and salt permeation has a lot of differences in a range is strongly recommended [15]. Somehow these days FO membranes conquer this obstacle with decreasing the thickness of the membrane by which is thinner than RO membranes. Furthermore, the problem of FO membranes can be determined as not being so stable under high pressure during PRO processes [16]. The comparison between RO, FO and PRO showed in Figure 2.

It is worth bearing in mind that PRO has several characteristics which should be compared with other industrialized analyses such as RO and FO [18].

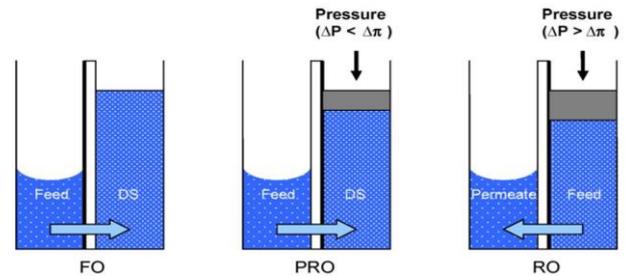

Figure 2: Comparison between RO, FO, and PRO according to $\Delta\pi$ and $\Delta P$ [17]

## II. EXPERIMENTAL

### A. PRO Process Design at Estuaries

One of the most common approaches for considering applications regarding the PRO process is to estimate the net energy which can be extracted at estuaries that rivers meet the sea. The amount of global river discharge is about 37,300 km$^3$ annually. If dilution of the river water is into the seawater, the potentiality to harvest energy is about 27,200 TW annually (which is equivalent to the continuing power output of 3.1 TW). This is a tremendous amount for energy annually which is feasible to be 20% larger than the global energy harvesting in 2012 [1].

### B. PRO for Power Generation and Desalination

Sidney Loeb is the first person that used the PRO process with two solutions of high concentration and low concentration to produce electricity. Through PRO, river water flows through the way of osmotic pressure or pass the semi-permeable membrane from low concentration to high concentration. To generate electricity from the turbine, the draw solution will be depressurized to modify hydraulic energy. The aim of Sidney Loeb is to check the availability of power production through PRO. To make sure that PRO can produce power, he used freshwater and RO brine as a feed solution and chose Dead-Sea water as the draw solution. Using RO brine for feed solution has two advantages; First, some resources in dry areas are rare like river water and the second one is during the RO brine usage will not deplete the evaporation rate and will refill it [5].

## III. RESULTS AND DISCUSSION

### A. PRO Membranes

The membrane works as a wall to filter or separate in processes. Membrane technology is a process that has a wide area to search scientifically and covers all the approaches of engineering. Permeable and semi-permeable membranes will allow the substances to pass through them for filtration or separation [6]. Loeb published a disappointing PRO performance when RO membranes were utilized in the PRO module [19]. Concentration polarization was criticized for the underperformance of the PRO process. Loeb and co-workers recognized that the performance of PRO can be developed by decreasing the membrane thickness, for example, by eliminating the support fabric layer of the commercial BWRO membrane [13]. The improvement of the osmotic characterization of the semi-permeable membrane is still one of the significant obstacles through PRO. One of the major issues is related to the low water flux that has consequently resulted in unpromising osmotic power. The desired characteristic of PRO-based on the TFC membrane should be of a highly hydrophilic and thin layer. The conventional membranes have high water flux with lower salt rejection properties. It increases the fouling in PRO operation which ultimately deteriorates the harvesting of energy in longer periods. PRO is an emerging membrane process used for producing power density and desalination applications. It is due to its characteristics that can facilitate, capture and convert the free Gibbs energy or mixing into the useful energy resources. Salinity gradient energy is a zero-emission and sustainable technology that can be practically applied worldwide. PRO is considered a very clean technology which is because of no byproduct formation and no $CO_2$ emission [18]. For instance, there is potentiality in energy resources in every estuary where the river water faces the seawater or in each desalination plants that release concentrated brines in the sea [20]. Thin-film nanocomposite (TFN) membranes are widely used in recent years for desalination applications. It is mainly due to the achievement of high flux without compromising the salt rejection. In this regard, nanomaterials such as metal oxides and carbon have gained more attention to improve membrane performance. It can improve the characteristic properties such as hydrophilicity and strength of the material. Hydrophilicity is the main required property in attaining the higher water flux [21]. The conventional TFC RO membranes are dense skin layer, which aids in high salt retention and lower flux. Conversely, the PRO membrane should be of thin skin for the achievement of high flux. PRO-based on the mechanism of the osmotic pressure gradient. Thus, the design of the TFN membrane is the prerequisite for achieving high-power density.

*B. PRO Membranes*

TFC membranes generally possess asymmetric porous support and a top selective skin. The chemistry used in the manufacture of the membrane was basically related to the interfacial polymerization of trimesoylchloride and m-phenylene diamine. The essential difference between reverse and a direct osmosis membrane is the following: a reverse osmosis membrane must withstand pressures of 60 or 80 bar and requires, therefore, a compaction resistant sub-structure. In RO water and salt are flowing in the same direction. The microporous support provides the mechanical strength, while the selective layer performs the separation. In the PRO process described here, the pressures are much lower, and salt and water are flowing in opposite directions. The thickness of the polyamide layer is in the range of hundred nanometers due to the self-terminating nature of the interfacial cross-linking reaction. The advantages of fabricating TFC membranes via interfacial polymerization are that the structure and properties of the substrate and the selective layer can be individually tailored and optimized to achieve desired permeability and salt rejection [22]. TFC has been widely used for PRO processes to generate power. Thin Film composites are advantageous during PRO processes. As an illustration, these beneficial characteristics can be mentioned have high water flux, mechanically robust and are vigorous to perform well in a wide range of pH. Otherwise, they have some disadvantages too. To illustrate, one of the most challenging obstacles is that they are not enough tolerant to face oxidants and chemicals that contain chlorine [23]. Fabrication of these membranes is possible in the laboratory which is in a similar way to fabricate RO membranes while the substrate and active layer will be prepared by phase inversion and can be done by interfacial polymerization [24]. Approximately, TFC membranes perform better for having higher water permeability than CTA which is shown in Table 1 [6].

Table 1: Water permeability through HTI CTA membranes and TFC membranes (*- All the TFC membranes are fabricated by MPD and TMC solution)

| Feed Solution | Draw Solution | Water flux $L/m^2h$ | Membrane orientation | Membrane type | References |
|---|---|---|---|---|---|
| DI water | 0.5M NaCl | 18.6 | FO | HTI CTA membrane | [25] |
| DI water | 1 M NaCl | 16.8 | FO | HTI CTA membrane | [17] |
| 50Mm NaCl | 4 M NaCl | 37.8 | PRO | HTI CTA membrane | [26] |
| 50Mm NaCl | 4 M NaCl | 27 | FO | HTI CTA membrane | [26] |
| Tap water | 0.5 M NaCl | 10.3 | FO | HTI CTA membrane | [27] |
| DI water | 0.5 M NaCl | 32.2 | PRO | TFC membrane* | [26] |
| DI water | 1.5 M NaCl | 18 | FO | TFC membrane* | [8] |
| DI water | 2 M NaCl | 47.5 | PRO | TFC membrane* | [26] |
| 0.5M NaCl | 2 M NaCl | 11.8 | FO | TFC membrane* | [26] |
| DI water | 2 M NaCl | 33 | PRO | TFC membrane* | [28] |
| 3.5% M NaCl | 2 M NaCl | 1 | PRO | TFC membrane* | [28] |

However, the salinity gradient resource has an influence on the water flux of the membrane. If the driving force through the membrane increases, then simultaneously the water flux enhances as well. To illustrate, if the water flux goes down from 33 $L/m^2h$ to 15 $L/m^2h$ then there is an enhancement for TDS (Total Dissolved Solids) of feed solution that will change from 0 to 3.5 % NaCl (Table 1) because of much more internal concentration polarization at higher feed solution

**Recent Advances of Nanomaterials in Membranes for Osmotic Energy Harvesting by Pressure Retarded Osmosis**

TDS. From Table 1, the results showed a higher water flux can be achieved when the membrane is operating on the PRO mode than on the FO mode. Furthermore, the results in Table 1 shows that the average water flux in the TFC membranes is higher than in the CTA membranes. Therefore, using the TFC membrane in the PRO process is desirable to increase the membrane flux and hence power density.

*C. Thin Film Nanocomposite (TFN) Membranes*

Thin film nanocomposite (TFN) membrane is a novel type of composite membranes developed via interfacial polymerization (IP) process. Nanoparticles are consolidated within the thin polyamide (PA) dense layer of the thin-film composite (TFC) membrane with the intention of enhancing the characteristics of the interracially polymerized layer. Whether nanocomposite membranes could achieve one or two of the performance enhancements, such as the balance of permeability and selectivity, and nanomaterials incorporated into polymeric membranes for water treatment. Firstly, 0D nanoparticles ($TiO_2$, $Al_2O_3$, $ZrO_2$, $SiO_2$, $ZnO$, $AgO$, etc.), consisting of metal oxides and metals, etc. are usually introduced into polyamide layer to enhance the surface hydrophilicity of membranes. Secondly, in addition to the hydrophilicity of nanoparticles, water transport channels of 1D nanotubes (single/multiwall carbon nanotubes, titanic nanotubes, etc.) also facilitate the permeability and fouling resistance of polymeric membranes. Finally, the 3D nanomaterials are porous according to their cage-like structure. For example, to improve the hydrophilicity and/or surface charge density, without reducing the separation efficiency of the TFC membrane.

Hydrophilicity is the chief determining parameter of the performance of TFN membranes. As a matter of fact, many types of inorganic nanomaterials have been incorporated with the polyamide layer. These nanoparticle fillers are commonly adopted to varieties of dimensional nanoscale (0D,1D,2D,3D) materials. As an illustration, 3D zeolites are microporous crystalline aluminosilicates with well-defined nanoscale pore structures in their regular frameworks. The hydrophilicity and permeability of 3D zeolites facilitate the permeability/selectivity and fouling resistance of ultrafiltration and reverse osmosis membranes [26]. Until now, 0D and 1D nanomaterials are prevalently used in the fabrication of TFN membranes. Works of literature on modification of TFN by 3D zeolites are sparsely reported. 3D Nanomaterials have an advantage as higher surface area, water absorption capacity and assembly on the polyamide layer. Thus, the current study is dedicated to investigating the zeolite incorporated in the TFN membrane for PRO application. Zeolite is a 3D nanomaterial which is highly hydrophilic as well as the salt rejection properties. It is mainly due to the cage-like structure of the zeolite. Table 2 shows the recent nanomaterials with the modification of other materials incorporated with the PRO membranes for increasing the main characteristics for boosting up the performance through PRO analyses.

TFN is referred to the incorporation of nanomaterials in the polyamide layer [39][40][41]. Recently, TFN membranes are widely used due to their potential to pass the obstacles such as the curve's trade-off among water permeability and salt rejection in comparison with the normal TFC membrane. [42][43]. Although there are impressive superior accomplishments, some challenges came across during the fabrication of the TFN membranes [44][45]. More research in this area is still needed to develop the TFN membrane with greater performance efficiency, reliability, and stability for industrial implementation. The method to improve TFN membrane fabrication can be achieved modification of hydrophilic nanofillers. These can help to develop the defect-free which is organic and the fabrication of nanofillers layer containing PA/inorganic. Subsequently, the usage and the selection of nanofillers in the fabrication of TFN membranes have to depend on the characteristics of the feed [39].

IV. CONCLUSION

PRO is one of the most promising applications for power generation and membranes are the heart of this method. In order to improve this process, there are a variety of nanoparticles that are somehow promising to be incorporated with composite membranes. In addition, the recent results have shown that they are more promising to use TFN membranes in the PRO process in comparison with the control TFC membrane. Improving the performance of TFN membranes through PRO process is mostly rely on the nanomaterials using in the substrate or active layer (polyamide layer). It has been observed that using nanomaterials in TFN membranes has led into having higher water flux and consequently to have a higher power density compared to the control TFC membrane. Working under different testing conditions will result in having various performance ranges which are between the 1.7 – 38.0 $W/m^2$. In addition, achieving better performance through the PRO process relies on the parameters such as different concentrations in the draw and feed solution and operation pressure. Therefore, some suggestions are recommended as the future directions of this research to enhance TFN and hollow fiber membranes for this application. In order to further improve the membrane flux and power density of the PRO system, the desired membrane should be of a very thin selective layer and the substrate layer. Thereby, high power density can be achieved in PRO techniques. The thickness and the structure of the substrate also play a critical role in determining the flux and power density. Hence it is recommended to fabricate a thin substrate to facilitate the transport of water.

**Table 2:** Recent nanomaterials incorporated with PRO membranes for PRO application process

| Membrane | Modification | Water Permeability (L m$^{-2}$ h$^{-1}$ bar$^{-1}$) | Water Flux (L m$^{-2}$ h$^{-1}$) | Power Density (Wm$^{-2}$) | Inference | Reference |
|---|---|---|---|---|---|---|
| TFC Hollow Fiber PES | GO | 1.36 | - | 14.6 | GO improved water permeability. | [29] |
| TFN membrane | Melamine-based Covalent Organic Framework (COF) | 3.87 | 42.5 | 12.1 | Melamine enhanced hydrophilicity and porosity. | [30] |
| TFN membrane | AgNP | 9.5 | 44.8 | - | "Phase inversion + LbL assembly" process developed water permeability and anti-biofouling property. | [31] |
| TFN membrane | CNT | 1.42 | 21.82 | 1.7 | TFN membrane takes advantage of the enhanced porosity and hydrophilicity induced by f-CNT. | [32] |
| TFN membrane | GO & HNT | 2.4 | - | 16.7 | The fabricated membrane developed power density and greater mechanical strength than commercial PRO membrane. | [33] |
| TFC membrane | The tiered structure of polyetherimide (PEI) reinforced by f-CNT | 3.04 | 23.2 | 17.3 | This membrane improved the mechanical properties along with the porosity of the support. | [34] |
| TFC membrane | Aminosilane | 2.23 | 36 | 12.8 | Grafting aminosilane culminated in having higher hydrophilicity, membrane strength of PRO membranes. | [35] |
| TFC Hollow Fiber membrane | Polyelectrolytes | 2 | 38 | 16.2 | The modified membrane attributed to the intrinsic properties, reducing S parameter due to hydrophilicity. | [36] |
| Hollow Fiber membrane | Polyamide-imide (PAI)/ glutaraldehyde (GA) | 0.66 | 27 | 4.3 | The hollow fiber membrane incorporated with PAI/GA has a superior antifouling property while the power density will remain constant. | [37] |
| TFC Hollow Fiber PES | CaCl$_2$ | 3.8 | 132 | 38 | TFC hollow fiber membrane incorporated with CaCl$_2$, has small pore size along with narrow distribution resulted in having strong support and highest water permeability. | [38] |

# Recent Advances of Nanomaterials in Membranes for Osmotic Energy Harvesting by Pressure Retarded Osmosis

**Arvin Shadravan** is a PhD student at the department of Industrial and Systems Engineering at Texas A&M University. He earned his BS and MS degree in Chemical Engineering.

**Mahmood Amani** is an associate professor at Texas A&M University at Qatar. He is the Petroleum Engineering Program Coordinator at Texas A&M University at Qatar, has broad experience in both industry and academia. He holds 2 U.S. Patents.